\documentclass[%
aip,
amsmath,amssymb,
reprint,%
]{revtex4-1}

\usepackage{graphicx}
\usepackage{geometry}
\usepackage{dcolumn}
\usepackage{bm}
\usepackage{xcolor}
\usepackage[T1]{fontenc}
\usepackage{mathptmx}
\usepackage{eqnarray}

\begin{document}
\preprint{AIP/123-QED}
\title{Four Wave-Mixing in a Microstrip Kinetic Inductance Travelling Wave Parametric Amplifier}
\author{Samuel Goldstein}
\email{samuel.goldstein@mail.huji.ac.il}
\author{Naftali Kirsh}%
\author{Elisha Svetitsky}%
\author{Yuval Zamir}%
\author{Ori Hachmo}%
\affiliation{ 
The Racah Institute of Physics, The Hebrew University of Jerusalem, Givat Ram Campus, Jerusalem, 9190401 Israel
}%
\author{Clovis Eduardo Mazzotti de Oliveira}
\affiliation{%
The Harvey M. Krueger Family Center for Nanoscience and Nanotechnology, The Hebrew University of Jerusalem, Givat Ram Campus, Jerusalem, 9190401 Israel
}%
\author{Nadav Katz}%
\affiliation{ 
The Racah Institute of Physics, The Hebrew University of Jerusalem, Givat Ram Campus, Jerusalem, 9190401 Israel
}%
\date{\today}

\begin{abstract}
Superconducting quantum circuits are typically operated at low temperatures (mK), necessitating cryogenic low-noise, wide-band amplifiers for signal readout ultimately also compatible with room temperature electronics. While existing implementations partly meet these criteria, they suffer from certain limitations, such as rippled transmission spectra or limited dynamic range, some of which are caused by the lack of proper impedance matching. We develop a MIcrostrip Kinetic Inductance Travelling Wave Amplifier (MI-KITWA), exploiting the nonlinear kinetic inductance of tungsten-silicide for wave-mixing of the signal and a pump, and engineer the impedance to $50 \Omega$, while decreasing the phase velocity, with benefit for the amplification. Despite losses, pumping on our device amplifies the signal by 15 dB over a 2 GHz bandwidth. 
\end{abstract}

\maketitle
The rapid advance of the last few years in the field of superconducting qubits has led to a demand for reliable cryogenic microwave amplifiers, characterized by low noise and large bandwidth\cite{krantz2019quantum}. Travelling wave parametric amplifiers present a solution\cite{ranzani2018kinetic} with two leading implementations: 1) Josephson Travelling Wave Parametric Amplifiers (JTWPA), that employ the Josephson Junction's nonlinear (tunable) inductance to yield a nonlinear wave-equation and wave-mixing\cite{sweeny1985travelling,o2014resonant,white2015traveling,macklin2015near}, and 2) Kinetic Inductance Travelling Wave Amplifiers (KITWA)\cite{eom2012wideband,eom2012wideband,bockstiegel2014development,adamyan2016superconducting,vissers2016low,chaudhuri2017broadband}, which exploit the high nonlinear kinetic inductance of certain materials (e.g. NbTiN or NbN).
Devices of both implementations share the challenge of matching their impedance $Z=\sqrt{(L_l/C_l)}$, where $L_l$ and $C_l$ are the inductance and capacitance per unit length, to the $50\Omega$ of conventional high-bandwidth electronics. The boosted $L_l$ increases $Z$ by up to an order of magnitude causing ripples in the transmission spectrum\cite{chaudhuri2017broadband,pozar2009microwave}. Various attempts to cope with this issue include adiabatic tapers at the beginning and end of the amplifier\cite{klopfenstein1956transmission}, shunting the transmission line by fractal structures\cite{adamyan2016superconducting}, and by multiple resonators\cite{white2015traveling} to lower $Z$.

\begin{figure}[b!]
\includegraphics[clip,trim=0.5cm 0cm 0cm 0cm,width=0.45\textwidth]{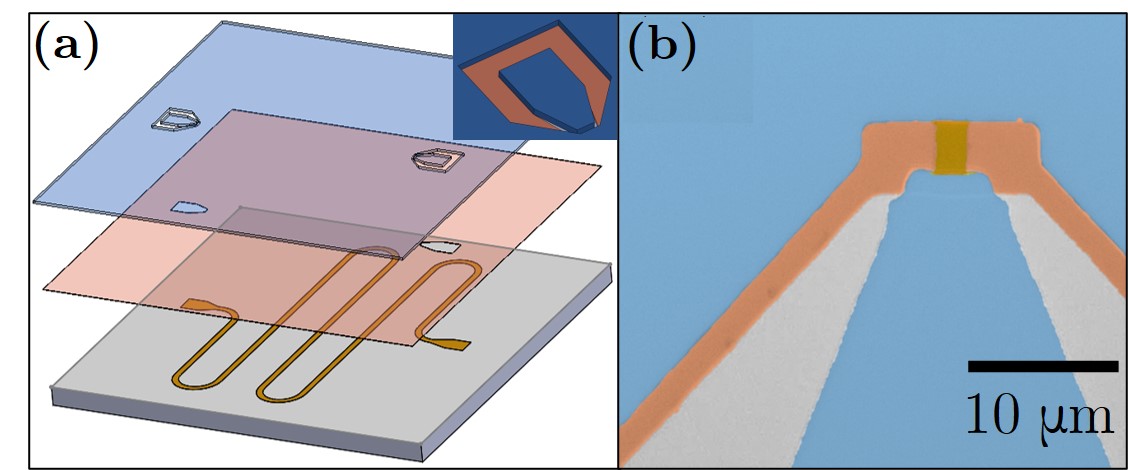}
\caption{\label{fig:Fabrication}
Fabrication of the MI-KITWA. (a) Graphic illustration of the layers: On the bulk Si substrate (grey), the WSi (brown) is patterned, followed by the dielectric (orange), and finally covered by the Al ground (blue). Inset: Launch pad for bonding with isolation from ground. (b) SEM image of launch pad with all layers visible (colors matching illustration).}
\end{figure}

\begin{figure*}[!]
\includegraphics[clip,trim=11cm 2cm 2cm 2cm,width=\textwidth]{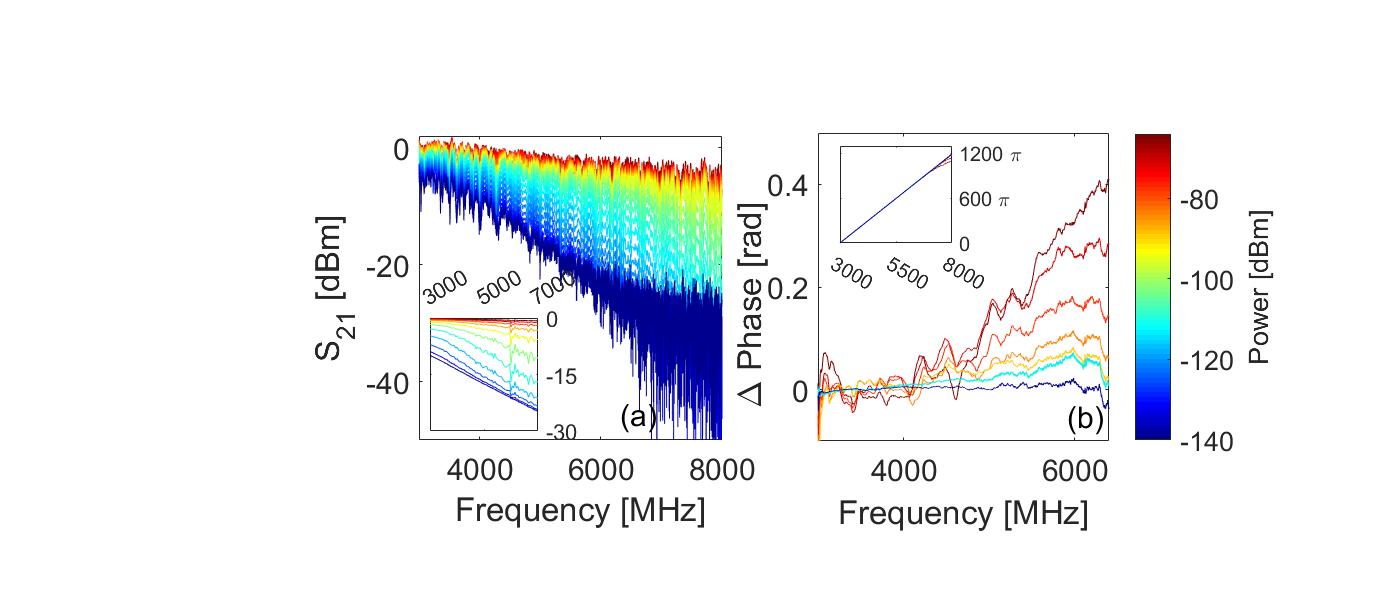}
\caption{\label{fig:SingleTone} Nonlinearity in single tone measurements. (a) Magnitude of measured signal transmission, as the input power is altered. Inset: Simulation, accounting also for loss saturation according to the standard TLS model\cite{kirsh2017revealing,phillips1987two} (assuming a low power loss tangent of $5\times10^{-4}$)\cite{o2008microwave}. (b) Unwrapped nonlinear phase of measured signal, referenced to the lowest power and to the individual power level's phase at $3 GHz$. Inset: Linear phase, relative to lowest measured frequency.}
\end{figure*}
While the various attempts to match the impedance partially succeed, they suffer from certain disadvantages: Adiabatic tapers extend the waveguide notably and despite the effort, ripples appear in the transmission spectrum. The JTWPA's primary problem is a lower dynamic range, and it too sometimes fails in fully matching the impedance of the external electronics\cite{eichler2014controlling}. The fractal structure is sensitive to fabrication errors and spurious ground plane resonances, requiring wire bonds across the trace\cite{adamyan2016superconducting}.

Here, we present a MIcrowave Kinetic Inductance Travelling Wave Amplifier (MI-KITWA). Instead of Nb alloys, we apply a nanometric layer of amorphous tungsten-silicide, WSi, whose kinetic inductance is sufficiently large, allowing us to neglect the magnetic inductance. Whereas implementations of other KITWAs we know of are coplanar waveguides with micronic gaps to the surrounding ground plates\cite{erickson2017theory}, the $C_l$ of the MI-KITWA resembles that of a parallel plate capacitor with the thickness of a few nm, and can be engineered rather easily to $50 \Omega$. 

A significant outcome of increasing $C_l$ is a drop in the phase velocity, $v_{ph}=1/\sqrt{(L_l C_l)}$, permitting us to shorten the amplifier by an order of magnitude, while still obtaining appreciable amplification. By fabrication of superconducting coplanar waveguide resonators in a separate experiment, we find that for the relevant cross section the kinetic inductance per unit length is $L_{k} \sim 50 \mu H/m$. Consequently we require $C_l \sim 20 nF/m$ to equate the MI-KITWA's impedance to $50\Omega$. These high values of $C_l$ and $L_l$ predict a theoretical $v_{ph} \sim 0.004c$, where $c$ is the speed of light in vacuum. We stress the implication of this result: The effective length of an $11.6 cm$ MI-KITWA extends to $8.7 m$ of an equivalent coplanar waveguide with $v_{ph}= 0.3c$. Even through a short trace ($11.6 cm$ here) the transmitted tones can accumulate a significant nonlinear phase sufficient for wave-mixing, and the MI-KITWA can then be much shorter than previous implementations\cite{eom2012wideband}. Concurrently, our simulations indicate that lower $v_{ph}$ enhances the amplification (while narrowing the operational bandwidth).

\begin{figure*}[!]
\includegraphics[clip,trim=1cm 11cm 1cm 11cm,width=\textwidth]{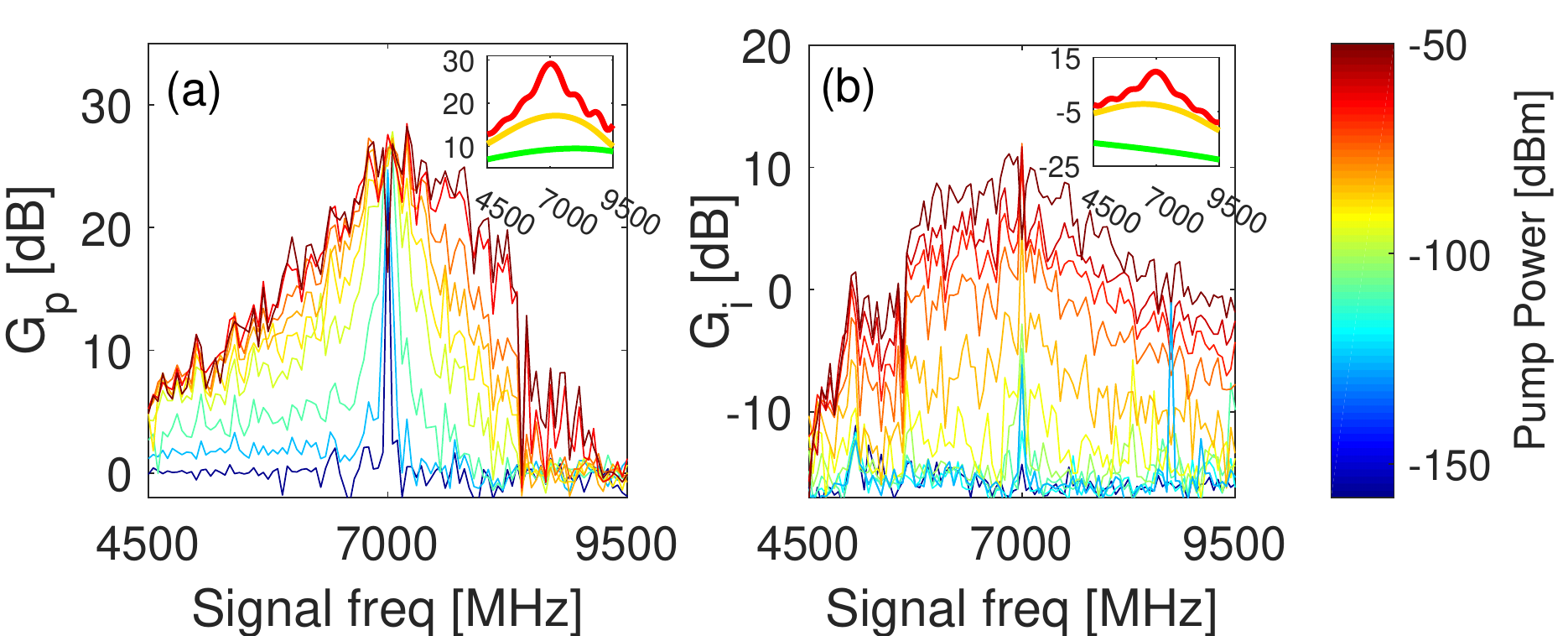}
\caption{\label{fig:SignalAndIdlerGain} Two-tone experiments demonstrate the parametric amplification and wave-mixing in the MI-KITWA. (a) Pump amplification factor $G_p$ for a signal at various $f_s$ (at $-140 dBm$), when the power of pump at $f_p=7.0 GHz$ is increased. Inset: Simulations for pump powers $-50 dBm$ (red), $-80 dBm$ (yellow), and $-110 dBm$ (green) show consistent amplitudes and bandwidth of $G_p$ as in experiment. (b) Idler conversion efficiency measured at $f_i=2f_p-f_s$. Inset: Simulation with color map as in (a).}
\end{figure*}

The nonlinear inductance is scaled by a current $I_{\star}$ (comparable to the critical current of the trace)  by\cite{eom2012wideband,erickson2017theory}:
\begin{equation}
L_{kin}=L_0 \Big(1+(I/I_{\star})^2 \Big)   
\end{equation}
The power dependence of $L_{kin}$ leads to a power-dependent wave-equation, which we here solve in the four-wave mixing ansatz (relevant to our experiment). In this case the current can be written as a linear combination of pump, signal, and idler tones ($I_p$, $I_s$, and $I_i$). Under the undepleted pump assumption, $|I_p|>>|I_s|,|I_i|$, the analytical solution for $I_p$ is\cite{eom2012wideband}:
\begin{equation}
I_p(z)=I_p(0)exp\Big(i(k_p z-\omega t)+ik_p z \gamma \Big) 
\label{eq:PumpSolution} 
\end{equation}
where $k_p=2\pi f_p/v_{ph}$ is the pump's wave number and $\gamma = |I_p(0)|^2/\big(2|I_{\star}|^2\big)$, i.e. in addition to the ordinary linear phase (inner brackets in exponent of Eq. \ref{eq:PumpSolution}), the current also accumulates a nonlinear phase shift\cite{white2015traveling}, $k_p z\gamma$ (here we neglect losses).

In implementations of coplanar KITWAs, wave-mixing is further controlled by phase matching of the wave numbers, $k_s$, $k_p$, and $k_i$, corresponding to the three current tones. This is accomplished by dispersion engineering, in particular by periodic perturbations of the trace widths (and thus its impedance), a feature that creates stop-bands in the transmission spectrum, which also serve to restrain shock wave generation\cite{erickson2017theory}. However, with the strong nonlinearity of our device and its low $v_{ph}$, we abandon the stop bands and keep the width (and thus the impedance) constant along the trace, a property which limits unwanted ripples in the transmission spectrum. 

Fabrication of the MI-KITWA involves three steps, each shaping a nanometric layer aligned with that below, illustrated in Fig. \ref{fig:Fabrication}. The first layer is the $5 nm$ thick highly inductive WSi trace, covered by a dielectric, amorphous Si film, a few nm thicker than the WSi to ensure step coverage, and engineered to match $50 \Omega$. The recipe is completed by a global Al ground plane (interrupted only for the launchers).

All experiments are performed at $20 mK$. Prior to amplification measurements, we demonstrate the performance of our device with linear and nonlinear single-tone characterization. Initially, we estimate $v_{ph}$ by broadcasting a pulse through the MI-KITWA, and in a different measurement through a parallel control channel. We compare the delay in arrival (not shown), and find $v_{ph}$ consistent with the theoretical expectation. In a separate measurement we determine the critical current of the WSi trace to be $I_c \simeq 0.06 mA$.

Next, we plot the nonlinear magnitude of the transmission (Fig. \ref{fig:SingleTone}a). The power-dependent loss can be associated with saturation of two-level systems (TLSs)\cite{kirsh2017revealing}, and is consistent with \textit{ab initio} simulations (where we use $tan\delta = 5\times10^{-4}$)\cite{o2008microwave}. These predict stronger amplification with only moderate improvements, such as higher nonlinearity (i.e. lowering $I_c$).

\begin{figure}[t]
\includegraphics[clip,trim=5.2cm 10.8cm 5.1cm 11cm,width=0.5\textwidth]{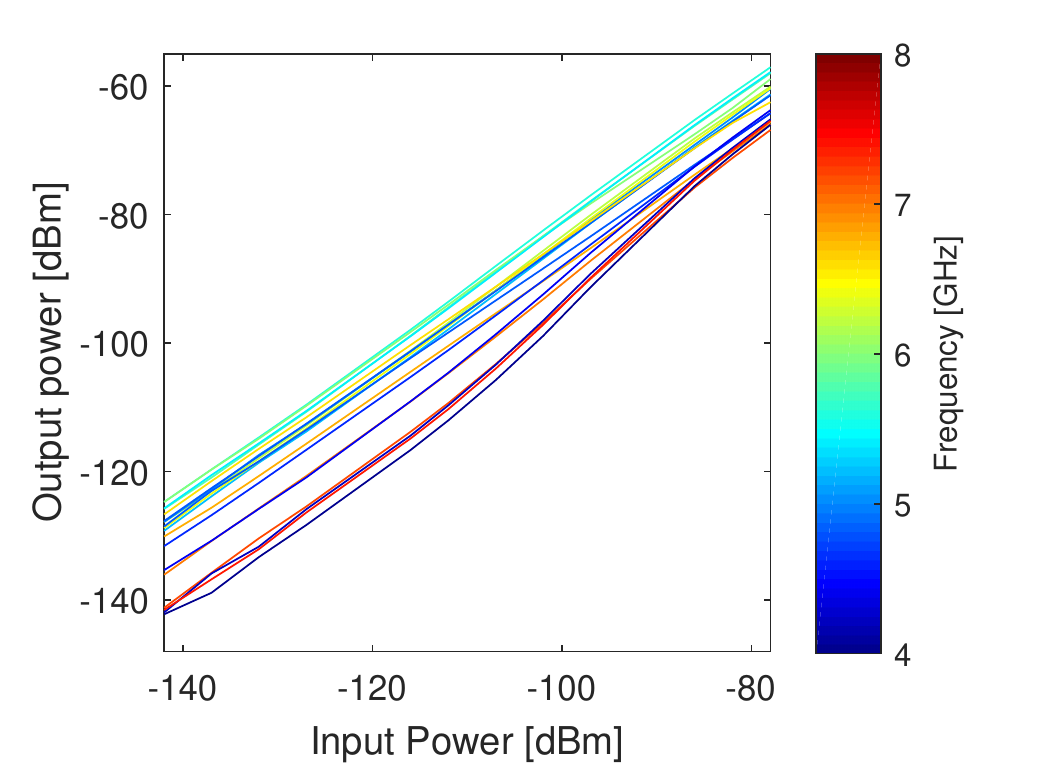}
\caption{\label{fig:DynamicNoise}Dynamic range: $G_p$ is nearly constant for various $f_s$ values, when we increase the signal power six orders of magnitude and keep the pump power $-53 dBm$ constant at $f_p=6.0 GHz$.}
\end{figure}

In turn, Fig. \ref{fig:SingleTone}b displays the \textit{nonlinear} phase accumulated through the device, allowing wave-mixing and amplification, cf. Eq. \ref{eq:PumpSolution}. This phase should be considered in light of the \textit{linear} phase accumulation of $\sim 10^3$ radians (Fig. \ref{fig:SingleTone}b. Inset). The linear increase of the slope of the nonlinear phase as a function of pump power follows Eq. \ref{eq:PumpSolution} (here shown up to 6.3 GHz beyond which losses make the extraction difficult at low power).

We now proceed to two-tone experiments, where wave-mixing between a low-powered signal at frequency $f_s$ and a high-powered pump at $f_p$ amplifies the former and produces an idler tone at $f_i$. The idler tone emerging at $f_i=2f_p-f_s$ is found as $f_p$ is kept constant at various powers, and the signal power is kept constant for changing $f_s$. Fig. \ref{fig:SignalAndIdlerGain}a shows the pump amplification factor
\begin{equation}
    G_p=\dfrac{|I_s^{pump}|^2}{|I_s^0|^2}
\end{equation}
where $I_s^{pump}(I_s^0)$ is the measured \textit{output} current at $f_s$  with(out) pump. We also plot the idler conversion efficiency\cite{stolen1982parametric}, i.e. the ratio between the measured idler and the signal input:
\begin{equation}
    G_i=\dfrac{|I_i^{out}|^2}{|I_s^{in}|^2}
\label{eq:IdlerConversion}
\end{equation}
These metrics are chosen instead of the more commonly used signal gain $G_s=|I_s^{out}|^2/|I_s^{in}|^2$ as they emphasize the nonlinearity of the MI-KITWA and the wave-mixing in it respectively\footnote{For our device, $G_s\simeq 10 dB$ over a $\sim 1.5 GHz$ bandwidth around the pump depending on the frequency.}. The strong idler (Fig. \ref{fig:SignalAndIdlerGain}b) confirms the wave-mixing taking place in the MI-KITWA.

An essential property of our amplifier is its large dynamic range (Fig. \ref{fig:DynamicNoise}). The transmission of the signal is nearly linear, as the  power increases by more than six orders of magnitude from $-140 dBm$ (below the single-photon equivalent power level), while a pump at $6 GHz$ is kept constant at $-53 dB$.

We conservatively estimate the MI-KITWA's noise temperature to be $0.25K <T_{eff}<0.41K$ by careful calibrations of the other components in our system, including amplifiers, attenuators, and  cables\footnote{The uncertainty arises from the losses which cannot be localized to certain temperature.}. This suggests the possibility of reaching the quantum noise limit with the MI-KITWA\cite{castellanos2008amplification,zobrist2019wide} by moderate improvement of the dielectric loss tangent and increasing the nonlinearity of our WSi (e.g. by different alloy concentration).

Applications of our device are not limited to amplification. The extraordinarily slow $v_{ph}$ allows the integration of the MI-KITWA or similar interferometric structures in superconducting quantum circuits, since the phase accumulation is at least an order of magnitude more compact than conventional coplanar structures. Furthermore, the joint distribution of signal and idler created in the nonlinear processes is expected to demonstrate two-mode squeezing\cite{guha2009gaussian}.

In summary, we have developed an impedance-matched microstrip amplifier, which derives its nonlinearity from WSi's power-dependent kinetic inductance. With a wide bandwidth, large dynamic range, and a sub-Kelvin noise temperature our amplifier benefits superconducting circuit readout protocols, especially with minor improvements lowering the dielectric loss. 

See the supplementary material for an expanded description of fabrication, the choice of pump frequencies, simulations including loss mechanisms, the experimental setup, and the noise temperature of our device.

We acknowledge the support of ISF grants 963.19 and 2323.19.

\bibliography{MIKITArefs}

\newpage
\onecolumngrid

	\begin{center}
		\textbf{Four Wave-Mixing in a Microstrip Kinetic Inductance Travelling Wave Parametric Amplifier}
	\end{center}
	\renewcommand{\thefigure}{S\arabic{figure}}
	\setcounter{figure}{0}
	\renewcommand{\theequation}{S\arabic{equation}}
	\setcounter{equation}{0} 
	
	\subsection{Fabrication}
	The MI-KITWA consists of three nanometric, mutually aligned layers as illustrated in Fig. 1 of the main text. The bottom layer is the $11.6 cm$ long trace itself, whose cross section is $2 \mu m \times 5 nm$ (during the research, other dimensions were tested, but here we present and discuss the results of the final version only), created by DC-magnetron sputtering a W-Si target (55\%/45\%) directly onto the high-resistivity Si substrate. This first layer is, like the two to follow, patterned by optical lithography and defined by selective wet-etch to form large pads at both ends of the trace. 
	
	The dielectric material for the second layer is chosen according to two properties: 1) its loss tangent $tan\delta $ and 2) its permittivity $\epsilon_r$. Materials with as low $tan\delta $ as possible are preferred to minimize dielectric loss, and $\epsilon_r$ regulates $C_l$ together with the thickness of this layer. We employ evaporated amorphous Si for this purpose, as its $tan\delta \sim 5\times 10^{-4}$, while $\epsilon_r$ is close to well-known values of $\epsilon_r^{Si}$ (depends weakly on morphology)\cite{o2008microwave,shur1989new}. Note that this lithographic step removes Si only from the WSi launchers (where metal in the next step enhances galvanic contact between wire-bonds and WSi pads). 
	
	The final layer of metal, by our choice evaporated Al, constitutes the ground, and the third and last lithography step now distinguishes between the all-covering Al and the aforementioned launch pads. Using a parallel (top) ground plate covering most of the chip ensures a common global ground, as opposed to the separated ground electrodes of co-planar wave guides, cut by the trace, and it also protects the trace from scratches during subsequent handling and packaging.
	\subsection{Gain and choice of pump frequency}
	In the main text we characterize amplification with the pump amplification factor $G_p$, defined there in Eq. 3, i.e. by normalizing the signal transmitted under presence of a pump with the signal transmitted, when no pump is applied. This metric emphasizes the nonlinearity of our device, and it allows ignoring loss from packaging and other external sources.
	A more conventional description of amplification is the signal gain:
	\begin{equation}
	G_s=\frac{|I_s^{out}|^2}{|I_s^{in}|^2} 
	\label{eq:Gain}
	\end{equation}
	i.e. $G_s$ normalizes by the input signal, and is thus smaller than $G_p$ due to internal loss mechanisms, which we simulate below.
	
	\begin{figure}[t]
		\includegraphics[clip,trim=0.6cm 11cm 0.1cm 11cm,width=0.9\textwidth]{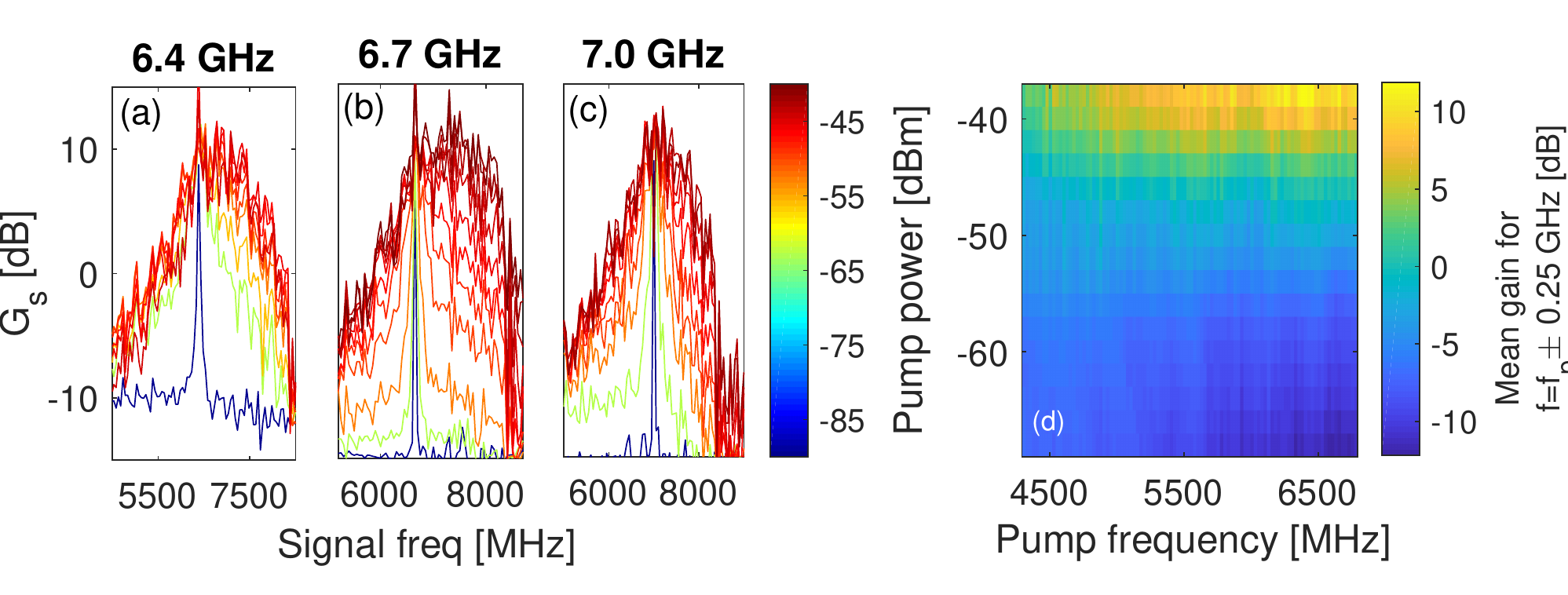}
		\caption{\label{fig:GainAndPump}\fontsize{10}{10}\linespread{1.5}\selectfont{} Signal gain (by the definition in Eq. \ref{eq:Gain}) for different pump powers, when the pump frequency $f_p$ is (a) 6.4 GHz, (b) 6.7 GHz, and (c) 7.0 GHz. Note the cutoff of our measurement line beyond 8.3 GHz due to circulator and HEMT characteristics. (d) The mean gain for all signal frequencies in a 250 MHz bandwidth around $f_p$.}
	\end{figure}
	
	$G_s$ is used to choose the optimal pump frequency for characterization experiments. In Fig. \ref{fig:GainAndPump}a-c we depict the gain around various pump frequencies, and Fig. \ref{fig:GainAndPump}d shows the mean gain for all signal frequencies around each $f_p$ for increasingly stronger pumps powers. Based on this, we operate our device only above 6000 MHz, where $G_s$ (and as a result also $G_p$) is enhanced. 
	
	\subsection{Theory and simulations}
	Following the established formalism of nonlinear optics, we simulate the performance of our amplifier by describing the propagating current as a linear combination of signal, pump, and idler tones\cite{eom2012wideband,white2015traveling}:
	
	\begin{equation}
	I=\frac{1}{2}\sum\limits_{n=s,p,i}I_n(z) e^{\big(i(k_n z-\omega t) \big)}+c.c.
	\label{eq:LinearCombination}
	\end{equation}
	where we ignore other wave-mixing processes\cite{erickson2017theory} than four wave-mixing with $2f_p=f_s+f_i$. The ansatz of Eq. \ref{eq:LinearCombination} is now used to solve the nonlinear wave equation
	\begin{equation}
	\frac{\partial^2I}{\partial z^2}=\frac{d}{dt}\Bigg(\frac{1}{\big(v_{ph}(I)\big)^2}\frac{dI}{dt}\Bigg)
	\end{equation}
	The nonlinearity of $v_{ph}$ stems from the power-dependent inductance (see Eq. 1 in the main text), and leads to the coupled equations, which under the undepleted pump assumption ($I_p>>I_s,I_i$) take the form
	
	\begin{subequations}
		\begin{equation}
		\frac{d I_p}{dz}=\frac{i k_p}{2 I_{\star}^2}|I_p|^2 I_p-\Gamma_p I_p
		\end{equation}
		\begin{equation}
		\frac{d I_p}{d_z}=\frac{ik_s}{2 I_{\star}^2}\Bigg( (|I_s|^2 I_p +I_i^*I_p^2e^{i\delta k z}\Bigg)-\Gamma_s I_s
		\end{equation}
		\begin{equation}
		\frac{d I_p}{d_z}=\frac{ik_i}{2 I_{\star}^2}\Bigg( (|I_i|^2 I_p +I_s^*I_p^2e^{i\delta k z}\Bigg)-\Gamma_i I_i
		\end{equation}
		\label{eq:CoupledMode}
	\end{subequations}
	
	Here $\delta k=2k_p-k_s-k_i$ is the mismatch between the wave numbers of three current tones in discussion, and as in the main text we include loss due to the dielectric material, expressed through the self-loss coefficient
	\begin{equation}
	\Gamma_p=\frac{2\pi f_p}{v_{ph}}\frac{1}{2}\frac{tan\delta_0}{\sqrt{(1+\Omega_{Rabi}^2T_1 T_2)}}
	\label{eq:Gamma_p}
	\end{equation}
	where $\Omega_{Rabi}$ is the Rabi frequency of the two-level-systems in the standard two-level-system (TLS) model\cite{kirsh2017revealing,phillips1987two}, $tan\delta_0$ is the loss tangent (without saturation effects), and  $T_1$ and $T_2$ are the decay and dephasing times\cite{pozar2009microwave,kirsh2017revealing}. Specifically, $\Omega_{Rabi}=2 d\cdot E/\hbar$, where $d$ is the dipole moment (in our simulations, $d=1 Db$). For both $T_1$ and $T_2$ we use the value $100 ns$.
	
	\begin{figure}[b!]
		\includegraphics[clip,trim=1.2cm 11cm 1.3cm 10.5cm,width=0.9\textwidth]{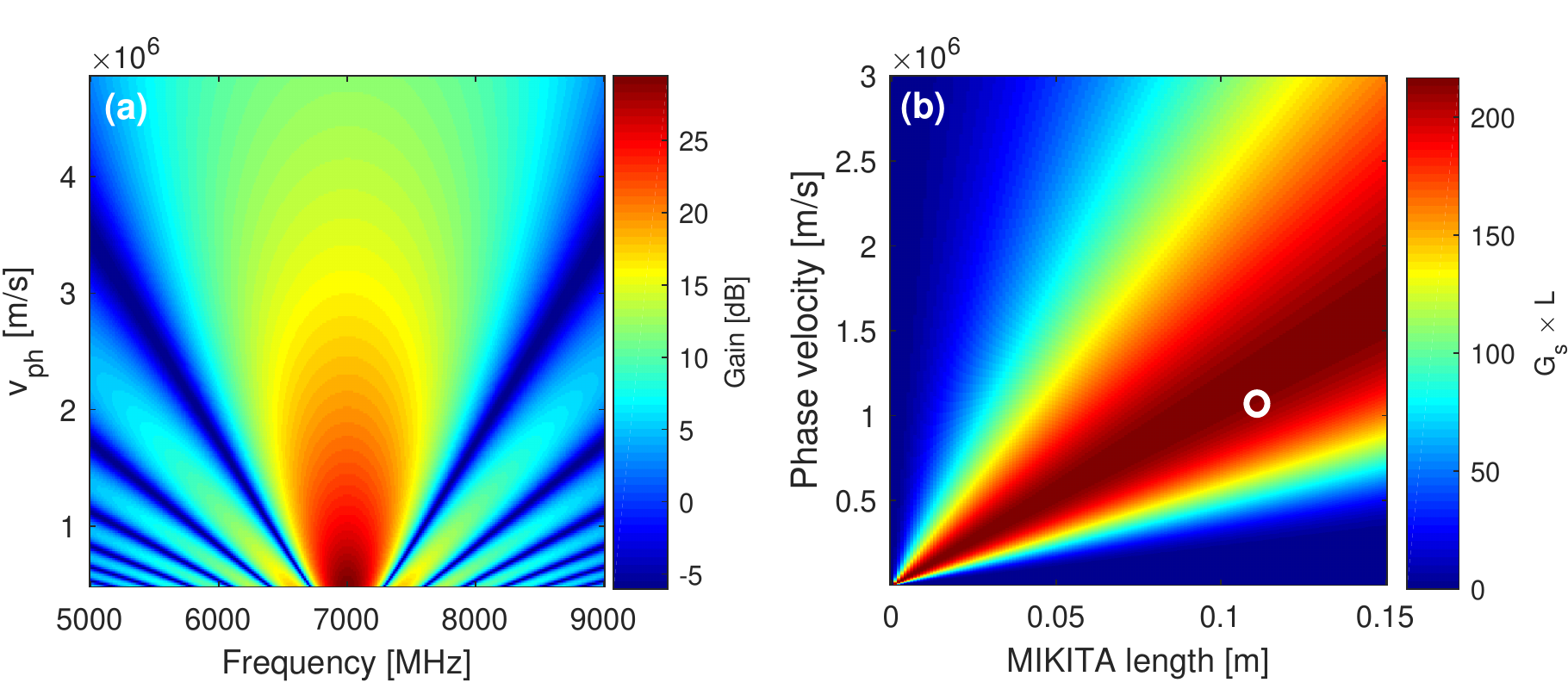}
		\caption{\label{fig:Vphase}\fontsize{10}{10}\linespread{1.5}\selectfont{} Simulations showing the impact of reducing the phase velocity on the amplification. (a) Amplification (in dB) for a $-48 dBm$ pump at $f_s=7 GHz$ and a signal of $-110 dBm$, neglecting nonlinear loss effects. (b) Signal gain ratio as defined in [\onlinecite{eom2012wideband}]: $G_s=\big(\gamma kz\big)^2+1$ times the power loss factor $L=e^{-2\Gamma_s^0 z}$, simulated without saturation effects for $f_s=7 GHz$ (with $\gamma=(I/I_{\star})^2=0.1$). The white circle indicates our working parameters $z=11.6 cm$ and $v_{ph}=0.0035 c$. Note that the theoretical amplification is a few $dB$ higher than what we found empirically, even after loss effects are taken into account. This discrepancy is likely due to the simulation's sensitivity to  parameters such as $tan\delta$ and $I_{\star}$.}
	\end{figure}
	
	In the TLS model, losses are due to absorption of energy by TLSs. At high pump powers the TLSs become saturated, and the energy absorption subsequently is decreased. The square-root in the denominator of Eq. \ref{eq:Gamma_p} accounts for absorption of a pump at $f_p$ not only by TLSs at the same frequency, but also those slightly detuned. When turning to two-tone measurements (with a strong pump at $f_p$ and a weaker signal at $f_s \neq f_p$), we assume that saturation effects are only caused by the pump, and that the absorption of the signal tone is linear. The signal loss is thus a function of the probability that TLSs at $f_s$ are excited by the signal itself rather than the pump. We approximate this probability by means of the saturation parameter of TLSs at $f_{s}$, when the pump is at $f_p$\cite{steck2007quantum}:
	
	\begin{equation*}
	s=\frac{\Omega_{Rabi}^2T_1T_2}{1+(\delta f)^2T_2^2}
	\end{equation*}
	where $\delta f=f_p-f_{TLS}$ (in this case $f_{TLS}=f_s$), and $\Omega_{Rabi}$ is the Rabi frequency of the pump. The excited population of the TLSs due to the pump is subsequently $\rho_{ee}=\tfrac{s}{2(1+s)}$. Thus we write both for the signal and the idler a pump dependent decay constant:
	\begin{equation*}
	\Gamma_{s,i}=\Gamma_{s,i}^{0}(1-2\rho_{ee})
	\end{equation*}
	with $\Gamma_{s,i}^{0}=\tfrac{1}{2}\tfrac{2\pi f_{s,i}}{v_{ph}}tan\delta_0$ being the linear decay constant, referring to the linear loss of the signal (idler) only, as if no pump were present.
		\begin{figure}[t!]
		\includegraphics[clip,trim=0cm 0cm 0cm 0cm, width=0.45\textwidth]{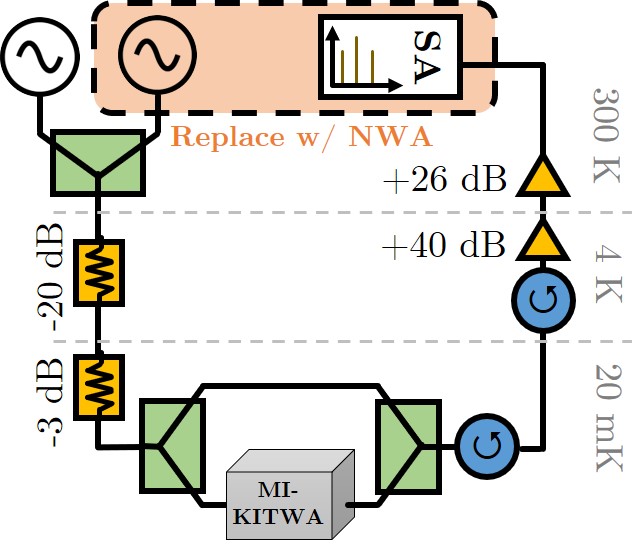}
		\caption{\label{fig:MI-KITWAsetup}\fontsize{10}{10}\linespread{1.5}\selectfont{} Experimental setup: Signal and pump are generated in two Anritsu MG3692B microwave generators (MG), merged in a power combiner (green) at room temperature, and passed through microwave attenuators (yellow). For calibration we connect the MI-KITWA in parallel with a control line through cold switches (purple) on both sides. The output passes through circulators (blue) to avoid reflections. Conventional amplifiers are placed at 4K (HEMT LNF-LNC) and $T_{room}$ (B\&Z LNA 2-8 $GHz$) before readout in a Keysight EXA N9010A spectrum analyzer (SA). In experiments without idler-measurements, the SA and one MG are replaced with an N5230A vector network analyzer (light orange).} 
	\end{figure}
	The analytic solution for the pump (Eq. \ref{eq:CoupledMode}a) is $I_p(z)=I_p(0)e^{ik_pz\tfrac{|I_p(0)|^2}{2 I_{\star}^2}-\Gamma_pz}$, as shown in Fig. 2b in the main text.
	
	Theoretical estimates of the amplification are now performed by initially simulating Eq. \ref{eq:CoupledMode}a to find the nonlinear transmission spectrum (see insert in Fig. 2a in main text) and the power-dependent $\Omega_{Rabi}$ of the pump throughout the MI-KITWA. Subsequently we numerically solve Eq. \ref{eq:CoupledMode}b and \ref{eq:CoupledMode}c. The resulting $G_P$ and $G_i$ for different pump powers appear in the main text's Fig. 3. 
	
	We also calculate the enhanced amplification by our device due to the very low $v_{ph}$. Fig. \ref{fig:Vphase}a visualizes how the amplification increases with decreasing $v_{ph}$, while the bandwidth is reduced. When expanding the length $z$ of the MI-KITWA, the loss intensifies as expressed by the loss factor $L=e^{-2 \Gamma_s z}$ (note the factor of 2, as this is power decay). Here we again ignore saturation effects. Also the signal gain $G_s=(\gamma k z)^2+1$ depends on both length and phase velocity\cite{white2015traveling}, and the product $G_s\times L$ thus predicts for which (${v_{ph},z}$) the highest amplification is accessible, as shown in Fig. \ref{fig:Vphase}b.

	
	\subsection{Noise temperature}
	
	Our device is the first component in a cascade of amplifiers (and effective attenuators) as shown in Fig. \ref{fig:MI-KITWAsetup}, and we find its noise temperature by solving the recursive equations for $N_{in}$ and $N_{out}$, the input and output noise of the components from the MI-KITWA and up through the line. Each component is treated either as an attenuator or as amplifier, in which case
	\begin{equation}
	N_{in}=\frac{N_{out}}{\tilde{G}_s}-k_BT_{eff}B
	\end{equation}
	where $B$ is the bandwidth (91 Hz in all our measurements), and $T_{eff}$ is the effective noise temperature, related to the noise figure $NF$, as $NF=10log_{10}\big(1+T_{eff}/T_0\big)$, where $T_0$ is the physical temperature. $\tilde{G}_s$ is the signal gain ratio (e.g. $G_s=20 dB \implies \tilde{G}_s=100$). 
	
	\begin{figure}[t]
		\includegraphics[clip,trim=4cm 10cm 4cm 10cm, width=0.7\textwidth]{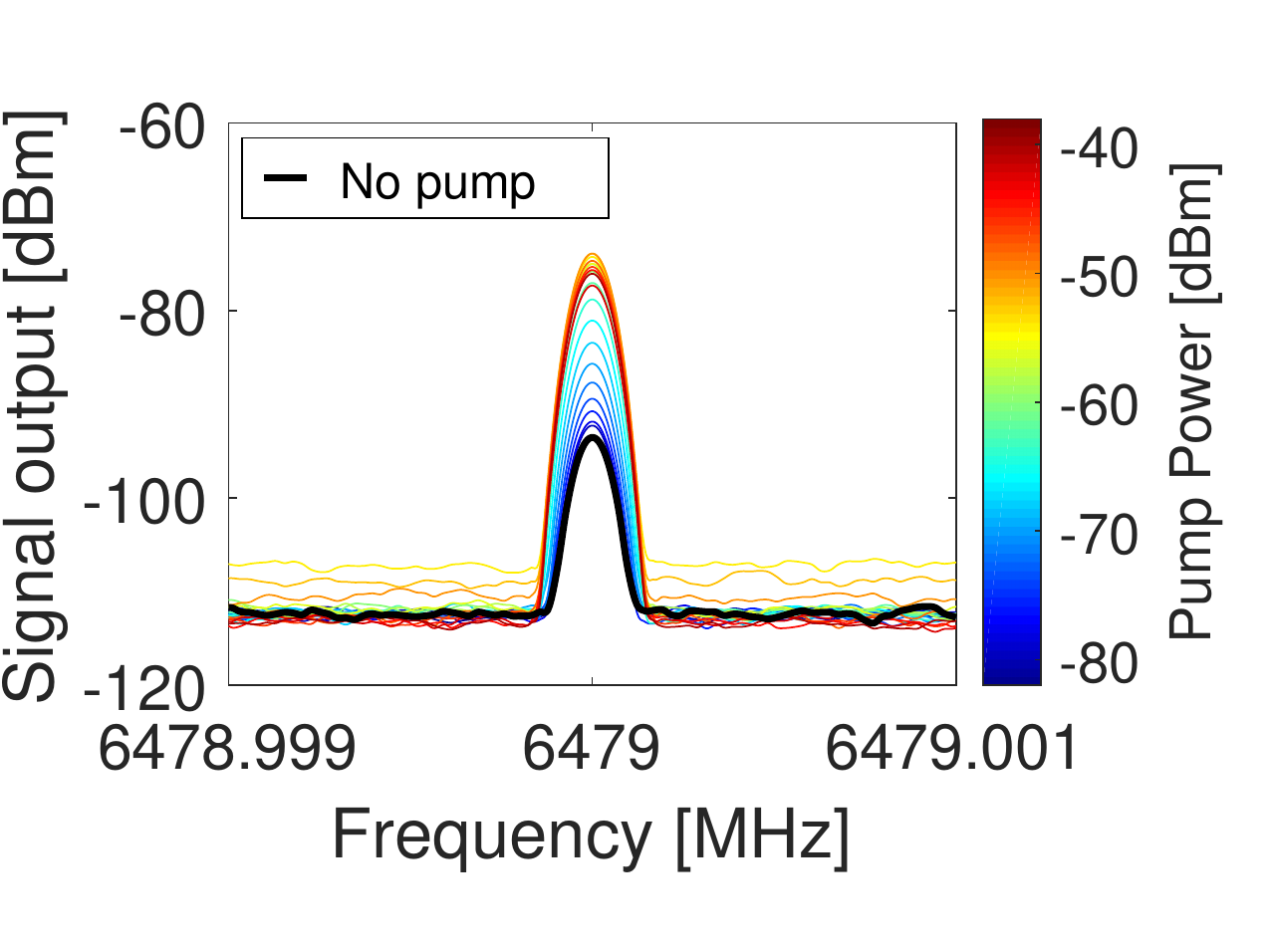}
		\caption{\label{fig:Noise}\fontsize{10}{10}\linespread{1.5}\selectfont{} Noise characterization. Signal output as measured after the cascade of switch, circulators, and commercial cryogenic amplifiers, as it appears in Fig. \ref{fig:MI-KITWAsetup}, when the signal is detuned by $100 MHz$ from the pump at $f_p=6379 MHz$.} 
	\end{figure}
	
	Attenuators have:
	\begin{equation}
	N_{in}=\frac{\Big(N_{out}+(\tilde{G}_s-1)k_BTB\Big)}{\tilde{G}_s}
	\end{equation}
	where $\tilde{G}_s$ still denotes the gain ($0<\tilde{G}_s<1$ for attenuators) analogue to the case of amplifiers.
	
	We determine the $G_s$ of the MI-KITWA by comparing the measured output in the case of the highest transmission on Fig. \ref{fig:Noise} with that from the control line (not shown), and we hereafter adopt the base line or \textit{noise floor}, $-108 dBm$, as the noise output at the end of the cascade. Rigorously solving for each component in the chain, including cable loss (treated as attenuation), sets the noise temperature of the MI-KITWA to be $0.25K <T_{eff}<0.41K$. The uncertainty results from an additional loss of $\sim 3 dB$ in the line measured, which we were unable to affiliate with localized components in the cascade (leaving the temperature and position of these sources of loss in the line uncertain).
	
	The $SNR$ of the measurement line with the MI-KITWA is read from Fig. \ref{fig:Noise} to be $15dB$ higher than the $SNR$ of the control measurement (with no pump).
	
	\bibliography{MIKITArefs}

\providecommand{\noopsort}[1]{}\providecommand{\singleletter}[1]{#1}%
\begin{thebibliography}{26}%
\makeatletter
\providecommand \@ifxundefined [1]{%
 \@ifx{#1\undefined}
}%
\providecommand \@ifnum [1]{%
 \ifnum #1\expandafter \@firstoftwo
 \else \expandafter \@secondoftwo
 \fi
}%
\providecommand \@ifx [1]{%
 \ifx #1\expandafter \@firstoftwo
 \else \expandafter \@secondoftwo
 \fi
}%
\providecommand \natexlab [1]{#1}%
\providecommand \enquote  [1]{``#1''}%
\providecommand \bibnamefont  [1]{#1}%
\providecommand \bibfnamefont [1]{#1}%
\providecommand \citenamefont [1]{#1}%
\providecommand \href@noop [0]{\@secondoftwo}%
\providecommand \href [0]{\begingroup \@sanitize@url \@href}%
\providecommand \@href[1]{\@@startlink{#1}\@@href}%
\providecommand \@@href[1]{\endgroup#1\@@endlink}%
\providecommand \@sanitize@url [0]{\catcode `\\12\catcode `\$12\catcode
  `\&12\catcode `\#12\catcode `\^12\catcode `\_12\catcode `\%12\relax}%
\providecommand \@@startlink[1]{}%
\providecommand \@@endlink[0]{}%
\providecommand \url  [0]{\begingroup\@sanitize@url \@url }%
\providecommand \@url [1]{\endgroup\@href {#1}{\urlprefix }}%
\providecommand \urlprefix  [0]{URL }%
\providecommand \Eprint [0]{\href }%
\providecommand \doibase [0]{http://dx.doi.org/}%
\providecommand \selectlanguage [0]{\@gobble}%
\providecommand \bibinfo  [0]{\@secondoftwo}%
\providecommand \bibfield  [0]{\@secondoftwo}%
\providecommand \translation [1]{[#1]}%
\providecommand \BibitemOpen [0]{}%
\providecommand \bibitemStop [0]{}%
\providecommand \bibitemNoStop [0]{.\EOS\space}%
\providecommand \EOS [0]{\spacefactor3000\relax}%
\providecommand \BibitemShut  [1]{\csname bibitem#1\endcsname}%
\let\auto@bib@innerbib\@empty
\bibitem [{\citenamefont {Krantz}\ \emph {et~al.}(2019)\citenamefont {Krantz},
  \citenamefont {Kjaergaard}, \citenamefont {Yan}, \citenamefont {Orlando},
  \citenamefont {Gustavsson},\ and\ \citenamefont
  {Oliver}}]{krantz2019quantum}%
  \BibitemOpen
  \bibfield  {author} {\bibinfo {author} {\bibfnamefont {P.}~\bibnamefont
  {Krantz}}, \bibinfo {author} {\bibfnamefont {M.}~\bibnamefont {Kjaergaard}},
  \bibinfo {author} {\bibfnamefont {F.}~\bibnamefont {Yan}}, \bibinfo {author}
  {\bibfnamefont {T.~P.}\ \bibnamefont {Orlando}}, \bibinfo {author}
  {\bibfnamefont {S.}~\bibnamefont {Gustavsson}}, \ and\ \bibinfo {author}
  {\bibfnamefont {W.~D.}\ \bibnamefont {Oliver}},\ }\bibfield  {title}
  {\enquote {\bibinfo {title} {A quantum engineer's guide to superconducting
  qubits},}\ }\href@noop {} {\bibfield  {journal} {\bibinfo  {journal} {Applied
  Physics Reviews}\ }\textbf {\bibinfo {volume} {6}},\ \bibinfo {pages}
  {021318} (\bibinfo {year} {2019})}\BibitemShut {NoStop}%
\bibitem [{\citenamefont {Ranzani}\ \emph {et~al.}(2018)\citenamefont
  {Ranzani}, \citenamefont {Bal}, \citenamefont {Fong}, \citenamefont
  {Ribeill}, \citenamefont {Wu}, \citenamefont {Long}, \citenamefont {Ku},
  \citenamefont {Erickson}, \citenamefont {Pappas},\ and\ \citenamefont
  {Ohki}}]{ranzani2018kinetic}%
  \BibitemOpen
  \bibfield  {author} {\bibinfo {author} {\bibfnamefont {L.}~\bibnamefont
  {Ranzani}}, \bibinfo {author} {\bibfnamefont {M.}~\bibnamefont {Bal}},
  \bibinfo {author} {\bibfnamefont {K.~C.}\ \bibnamefont {Fong}}, \bibinfo
  {author} {\bibfnamefont {G.}~\bibnamefont {Ribeill}}, \bibinfo {author}
  {\bibfnamefont {X.}~\bibnamefont {Wu}}, \bibinfo {author} {\bibfnamefont
  {J.}~\bibnamefont {Long}}, \bibinfo {author} {\bibfnamefont {H.-S.}\
  \bibnamefont {Ku}}, \bibinfo {author} {\bibfnamefont {R.~P.}\ \bibnamefont
  {Erickson}}, \bibinfo {author} {\bibfnamefont {D.}~\bibnamefont {Pappas}}, \
  and\ \bibinfo {author} {\bibfnamefont {T.~A.}\ \bibnamefont {Ohki}},\
  }\bibfield  {title} {\enquote {\bibinfo {title} {Kinetic inductance
  traveling-wave amplifiers for multiplexed qubit readout},}\ }\href@noop {}
  {\bibfield  {journal} {\bibinfo  {journal} {Applied Physics Letters}\
  }\textbf {\bibinfo {volume} {113}},\ \bibinfo {pages} {242602} (\bibinfo
  {year} {2018})}\BibitemShut {NoStop}%
\bibitem [{\citenamefont {Sweeny}\ and\ \citenamefont
  {Mahler}(1985)}]{sweeny1985travelling}%
  \BibitemOpen
  \bibfield  {author} {\bibinfo {author} {\bibfnamefont {M.}~\bibnamefont
  {Sweeny}}\ and\ \bibinfo {author} {\bibfnamefont {R.}~\bibnamefont
  {Mahler}},\ }\bibfield  {title} {\enquote {\bibinfo {title} {A
  travelling-wave parametric amplifier utilizing josephson junctions},}\
  }\href@noop {} {\bibfield  {journal} {\bibinfo  {journal} {IEEE Transactions
  on Magnetics}\ }\textbf {\bibinfo {volume} {21}},\ \bibinfo {pages}
  {654--655} (\bibinfo {year} {1985})}\BibitemShut {NoStop}%
\bibitem [{\citenamefont {O’Brien}\ \emph {et~al.}(2014)\citenamefont
  {O’Brien}, \citenamefont {Macklin}, \citenamefont {Siddiqi},\ and\
  \citenamefont {Zhang}}]{o2014resonant}%
  \BibitemOpen
  \bibfield  {author} {\bibinfo {author} {\bibfnamefont {K.}~\bibnamefont
  {O’Brien}}, \bibinfo {author} {\bibfnamefont {C.}~\bibnamefont {Macklin}},
  \bibinfo {author} {\bibfnamefont {I.}~\bibnamefont {Siddiqi}}, \ and\
  \bibinfo {author} {\bibfnamefont {X.}~\bibnamefont {Zhang}},\ }\bibfield
  {title} {\enquote {\bibinfo {title} {Resonant phase matching of josephson
  junction traveling wave parametric amplifiers},}\ }\href@noop {} {\bibfield
  {journal} {\bibinfo  {journal} {Physical review letters}\ }\textbf {\bibinfo
  {volume} {113}},\ \bibinfo {pages} {157001} (\bibinfo {year}
  {2014})}\BibitemShut {NoStop}%
\bibitem [{\citenamefont {White}\ \emph {et~al.}(2015)\citenamefont {White},
  \citenamefont {Mutus}, \citenamefont {Hoi}, \citenamefont {Barends},
  \citenamefont {Campbell}, \citenamefont {Chen}, \citenamefont {Chen},
  \citenamefont {Chiaro}, \citenamefont {Dunsworth}, \citenamefont {Jeffrey}
  \emph {et~al.}}]{white2015traveling}%
  \BibitemOpen
  \bibfield  {author} {\bibinfo {author} {\bibfnamefont {T.}~\bibnamefont
  {White}}, \bibinfo {author} {\bibfnamefont {J.}~\bibnamefont {Mutus}},
  \bibinfo {author} {\bibfnamefont {I.-C.}\ \bibnamefont {Hoi}}, \bibinfo
  {author} {\bibfnamefont {R.}~\bibnamefont {Barends}}, \bibinfo {author}
  {\bibfnamefont {B.}~\bibnamefont {Campbell}}, \bibinfo {author}
  {\bibfnamefont {Y.}~\bibnamefont {Chen}}, \bibinfo {author} {\bibfnamefont
  {Z.}~\bibnamefont {Chen}}, \bibinfo {author} {\bibfnamefont {B.}~\bibnamefont
  {Chiaro}}, \bibinfo {author} {\bibfnamefont {A.}~\bibnamefont {Dunsworth}},
  \bibinfo {author} {\bibfnamefont {E.}~\bibnamefont {Jeffrey}},  \emph
  {et~al.},\ }\bibfield  {title} {\enquote {\bibinfo {title} {Traveling wave
  parametric amplifier with josephson junctions using minimal resonator phase
  matching},}\ }\href@noop {} {\bibfield  {journal} {\bibinfo  {journal}
  {Applied Physics Letters}\ }\textbf {\bibinfo {volume} {106}},\ \bibinfo
  {pages} {242601} (\bibinfo {year} {2015})}\BibitemShut {NoStop}%
\bibitem [{\citenamefont {Macklin}\ \emph {et~al.}(2015)\citenamefont
  {Macklin}, \citenamefont {O’Brien}, \citenamefont {Hover}, \citenamefont
  {Schwartz}, \citenamefont {Bolkhovsky}, \citenamefont {Zhang}, \citenamefont
  {Oliver},\ and\ \citenamefont {Siddiqi}}]{macklin2015near}%
  \BibitemOpen
  \bibfield  {author} {\bibinfo {author} {\bibfnamefont {C.}~\bibnamefont
  {Macklin}}, \bibinfo {author} {\bibfnamefont {K.}~\bibnamefont {O’Brien}},
  \bibinfo {author} {\bibfnamefont {D.}~\bibnamefont {Hover}}, \bibinfo
  {author} {\bibfnamefont {M.}~\bibnamefont {Schwartz}}, \bibinfo {author}
  {\bibfnamefont {V.}~\bibnamefont {Bolkhovsky}}, \bibinfo {author}
  {\bibfnamefont {X.}~\bibnamefont {Zhang}}, \bibinfo {author} {\bibfnamefont
  {W.}~\bibnamefont {Oliver}}, \ and\ \bibinfo {author} {\bibfnamefont
  {I.}~\bibnamefont {Siddiqi}},\ }\bibfield  {title} {\enquote {\bibinfo
  {title} {A near--quantum-limited josephson traveling-wave parametric
  amplifier},}\ }\href@noop {} {\bibfield  {journal} {\bibinfo  {journal}
  {Science}\ }\textbf {\bibinfo {volume} {350}},\ \bibinfo {pages} {307--310}
  (\bibinfo {year} {2015})}\BibitemShut {NoStop}%
\bibitem [{\citenamefont {Eom}\ \emph {et~al.}(2012)\citenamefont {Eom},
  \citenamefont {Day}, \citenamefont {LeDuc},\ and\ \citenamefont
  {Zmuidzinas}}]{eom2012wideband}%
  \BibitemOpen
  \bibfield  {author} {\bibinfo {author} {\bibfnamefont {B.~H.}\ \bibnamefont
  {Eom}}, \bibinfo {author} {\bibfnamefont {P.~K.}\ \bibnamefont {Day}},
  \bibinfo {author} {\bibfnamefont {H.~G.}\ \bibnamefont {LeDuc}}, \ and\
  \bibinfo {author} {\bibfnamefont {J.}~\bibnamefont {Zmuidzinas}},\ }\bibfield
   {title} {\enquote {\bibinfo {title} {A wideband, low-noise superconducting
  amplifier with high dynamic range},}\ }\href@noop {} {\bibfield  {journal}
  {\bibinfo  {journal} {Nature Physics}\ }\textbf {\bibinfo {volume} {8}},\
  \bibinfo {pages} {623} (\bibinfo {year} {2012})}\BibitemShut {NoStop}%
\bibitem [{\citenamefont {Bockstiegel}\ \emph {et~al.}(2014)\citenamefont
  {Bockstiegel}, \citenamefont {Gao}, \citenamefont {Vissers}, \citenamefont
  {Sandberg}, \citenamefont {Chaudhuri}, \citenamefont {Sanders}, \citenamefont
  {Vale}, \citenamefont {Irwin},\ and\ \citenamefont
  {Pappas}}]{bockstiegel2014development}%
  \BibitemOpen
  \bibfield  {author} {\bibinfo {author} {\bibfnamefont {C.}~\bibnamefont
  {Bockstiegel}}, \bibinfo {author} {\bibfnamefont {J.}~\bibnamefont {Gao}},
  \bibinfo {author} {\bibfnamefont {M.}~\bibnamefont {Vissers}}, \bibinfo
  {author} {\bibfnamefont {M.}~\bibnamefont {Sandberg}}, \bibinfo {author}
  {\bibfnamefont {S.}~\bibnamefont {Chaudhuri}}, \bibinfo {author}
  {\bibfnamefont {A.}~\bibnamefont {Sanders}}, \bibinfo {author} {\bibfnamefont
  {L.}~\bibnamefont {Vale}}, \bibinfo {author} {\bibfnamefont {K.}~\bibnamefont
  {Irwin}}, \ and\ \bibinfo {author} {\bibfnamefont {D.}~\bibnamefont
  {Pappas}},\ }\bibfield  {title} {\enquote {\bibinfo {title} {Development of a
  broadband nbtin traveling wave parametric amplifier for mkid readout},}\
  }\href@noop {} {\bibfield  {journal} {\bibinfo  {journal} {Journal of Low
  Temperature Physics}\ }\textbf {\bibinfo {volume} {176}},\ \bibinfo {pages}
  {476--482} (\bibinfo {year} {2014})}\BibitemShut {NoStop}%
\bibitem [{\citenamefont {Adamyan}\ \emph {et~al.}(2016)\citenamefont
  {Adamyan}, \citenamefont {De~Graaf}, \citenamefont {Kubatkin},\ and\
  \citenamefont {Danilov}}]{adamyan2016superconducting}%
  \BibitemOpen
  \bibfield  {author} {\bibinfo {author} {\bibfnamefont {A.}~\bibnamefont
  {Adamyan}}, \bibinfo {author} {\bibfnamefont {S.}~\bibnamefont {De~Graaf}},
  \bibinfo {author} {\bibfnamefont {S.}~\bibnamefont {Kubatkin}}, \ and\
  \bibinfo {author} {\bibfnamefont {A.}~\bibnamefont {Danilov}},\ }\bibfield
  {title} {\enquote {\bibinfo {title} {Superconducting microwave parametric
  amplifier based on a quasi-fractal slow propagation line},}\ }\href@noop {}
  {\bibfield  {journal} {\bibinfo  {journal} {Journal of Applied Physics}\
  }\textbf {\bibinfo {volume} {119}},\ \bibinfo {pages} {083901} (\bibinfo
  {year} {2016})}\BibitemShut {NoStop}%
\bibitem [{\citenamefont {Vissers}\ \emph {et~al.}(2016)\citenamefont
  {Vissers}, \citenamefont {Erickson}, \citenamefont {Ku}, \citenamefont
  {Vale}, \citenamefont {Wu}, \citenamefont {Hilton},\ and\ \citenamefont
  {Pappas}}]{vissers2016low}%
  \BibitemOpen
  \bibfield  {author} {\bibinfo {author} {\bibfnamefont {M.~R.}\ \bibnamefont
  {Vissers}}, \bibinfo {author} {\bibfnamefont {R.~P.}\ \bibnamefont
  {Erickson}}, \bibinfo {author} {\bibfnamefont {H.-S.}\ \bibnamefont {Ku}},
  \bibinfo {author} {\bibfnamefont {L.}~\bibnamefont {Vale}}, \bibinfo {author}
  {\bibfnamefont {X.}~\bibnamefont {Wu}}, \bibinfo {author} {\bibfnamefont
  {G.}~\bibnamefont {Hilton}}, \ and\ \bibinfo {author} {\bibfnamefont {D.~P.}\
  \bibnamefont {Pappas}},\ }\bibfield  {title} {\enquote {\bibinfo {title}
  {Low-noise kinetic inductance traveling-wave amplifier using three-wave
  mixing},}\ }\href@noop {} {\bibfield  {journal} {\bibinfo  {journal} {Applied
  physics letters}\ }\textbf {\bibinfo {volume} {108}},\ \bibinfo {pages}
  {012601} (\bibinfo {year} {2016})}\BibitemShut {NoStop}%
\bibitem [{\citenamefont {Chaudhuri}\ \emph {et~al.}(2017)\citenamefont
  {Chaudhuri}, \citenamefont {Li}, \citenamefont {Irwin}, \citenamefont
  {Bockstiegel}, \citenamefont {Hubmayr}, \citenamefont {Ullom}, \citenamefont
  {Vissers},\ and\ \citenamefont {Gao}}]{chaudhuri2017broadband}%
  \BibitemOpen
  \bibfield  {author} {\bibinfo {author} {\bibfnamefont {S.}~\bibnamefont
  {Chaudhuri}}, \bibinfo {author} {\bibfnamefont {D.}~\bibnamefont {Li}},
  \bibinfo {author} {\bibfnamefont {K.}~\bibnamefont {Irwin}}, \bibinfo
  {author} {\bibfnamefont {C.}~\bibnamefont {Bockstiegel}}, \bibinfo {author}
  {\bibfnamefont {J.}~\bibnamefont {Hubmayr}}, \bibinfo {author} {\bibfnamefont
  {J.}~\bibnamefont {Ullom}}, \bibinfo {author} {\bibfnamefont
  {M.}~\bibnamefont {Vissers}}, \ and\ \bibinfo {author} {\bibfnamefont
  {J.}~\bibnamefont {Gao}},\ }\bibfield  {title} {\enquote {\bibinfo {title}
  {Broadband parametric amplifiers based on nonlinear kinetic inductance
  artificial transmission lines},}\ }\href@noop {} {\bibfield  {journal}
  {\bibinfo  {journal} {Applied Physics Letters}\ }\textbf {\bibinfo {volume}
  {110}},\ \bibinfo {pages} {152601} (\bibinfo {year} {2017})}\BibitemShut
  {NoStop}%
\bibitem [{\citenamefont {Pozar}(2009)}]{pozar2009microwave}%
  \BibitemOpen
  \bibfield  {author} {\bibinfo {author} {\bibfnamefont {D.~M.}\ \bibnamefont
  {Pozar}},\ }\href@noop {} {\emph {\bibinfo {title} {Microwave engineering}}}\
  (\bibinfo  {publisher} {John Wiley \& Sons},\ \bibinfo {year}
  {2009})\BibitemShut {NoStop}%
\bibitem [{\citenamefont {Klopfenstein}(1956)}]{klopfenstein1956transmission}%
  \BibitemOpen
  \bibfield  {author} {\bibinfo {author} {\bibfnamefont {R.~W.}\ \bibnamefont
  {Klopfenstein}},\ }\bibfield  {title} {\enquote {\bibinfo {title} {A
  transmission line taper of improved design},}\ }\href@noop {} {\bibfield
  {journal} {\bibinfo  {journal} {Proceedings of the IRE}\ }\textbf {\bibinfo
  {volume} {44}},\ \bibinfo {pages} {31--35} (\bibinfo {year}
  {1956})}\BibitemShut {NoStop}%
\bibitem [{\citenamefont {Kirsh}\ \emph {et~al.}(2017)\citenamefont {Kirsh},
  \citenamefont {Svetitsky}, \citenamefont {Burin}, \citenamefont {Schechter},\
  and\ \citenamefont {Katz}}]{kirsh2017revealing}%
  \BibitemOpen
  \bibfield  {author} {\bibinfo {author} {\bibfnamefont {N.}~\bibnamefont
  {Kirsh}}, \bibinfo {author} {\bibfnamefont {E.}~\bibnamefont {Svetitsky}},
  \bibinfo {author} {\bibfnamefont {A.~L.}\ \bibnamefont {Burin}}, \bibinfo
  {author} {\bibfnamefont {M.}~\bibnamefont {Schechter}}, \ and\ \bibinfo
  {author} {\bibfnamefont {N.}~\bibnamefont {Katz}},\ }\bibfield  {title}
  {\enquote {\bibinfo {title} {Revealing the nonlinear response of a tunneling
  two-level system ensemble using coupled modes},}\ }\href@noop {} {\bibfield
  {journal} {\bibinfo  {journal} {Physical Review Materials}\ }\textbf
  {\bibinfo {volume} {1}},\ \bibinfo {pages} {012601} (\bibinfo {year}
  {2017})}\BibitemShut {NoStop}%
\bibitem [{\citenamefont {Phillips}(1987)}]{phillips1987two}%
  \BibitemOpen
  \bibfield  {author} {\bibinfo {author} {\bibfnamefont {W.}~\bibnamefont
  {Phillips}},\ }\bibfield  {title} {\enquote {\bibinfo {title} {Two-level
  states in glasses},}\ }\href@noop {} {\bibfield  {journal} {\bibinfo
  {journal} {Reports on Progress in Physics}\ }\textbf {\bibinfo {volume}
  {50}},\ \bibinfo {pages} {1657} (\bibinfo {year} {1987})}\BibitemShut
  {NoStop}%
\bibitem [{\citenamefont {O’Connell}\ \emph {et~al.}(2008)\citenamefont
  {O’Connell}, \citenamefont {Ansmann}, \citenamefont {Bialczak},
  \citenamefont {Hofheinz}, \citenamefont {Katz}, \citenamefont {Lucero},
  \citenamefont {McKenney}, \citenamefont {Neeley}, \citenamefont {Wang},
  \citenamefont {Weig} \emph {et~al.}}]{o2008microwave}%
  \BibitemOpen
  \bibfield  {author} {\bibinfo {author} {\bibfnamefont {A.~D.}\ \bibnamefont
  {O’Connell}}, \bibinfo {author} {\bibfnamefont {M.}~\bibnamefont
  {Ansmann}}, \bibinfo {author} {\bibfnamefont {R.~C.}\ \bibnamefont
  {Bialczak}}, \bibinfo {author} {\bibfnamefont {M.}~\bibnamefont {Hofheinz}},
  \bibinfo {author} {\bibfnamefont {N.}~\bibnamefont {Katz}}, \bibinfo {author}
  {\bibfnamefont {E.}~\bibnamefont {Lucero}}, \bibinfo {author} {\bibfnamefont
  {C.}~\bibnamefont {McKenney}}, \bibinfo {author} {\bibfnamefont
  {M.}~\bibnamefont {Neeley}}, \bibinfo {author} {\bibfnamefont
  {H.}~\bibnamefont {Wang}}, \bibinfo {author} {\bibfnamefont {E.~M.}\
  \bibnamefont {Weig}},  \emph {et~al.},\ }\bibfield  {title} {\enquote
  {\bibinfo {title} {Microwave dielectric loss at single photon energies and
  millikelvin temperatures},}\ }\href@noop {} {\bibfield  {journal} {\bibinfo
  {journal} {Applied Physics Letters}\ }\textbf {\bibinfo {volume} {92}},\
  \bibinfo {pages} {112903} (\bibinfo {year} {2008})}\BibitemShut {NoStop}%
\bibitem [{\citenamefont {Eichler}\ and\ \citenamefont
  {Wallraff}(2014)}]{eichler2014controlling}%
  \BibitemOpen
  \bibfield  {author} {\bibinfo {author} {\bibfnamefont {C.}~\bibnamefont
  {Eichler}}\ and\ \bibinfo {author} {\bibfnamefont {A.}~\bibnamefont
  {Wallraff}},\ }\bibfield  {title} {\enquote {\bibinfo {title} {Controlling
  the dynamic range of a josephson parametric amplifier},}\ }\href@noop {}
  {\bibfield  {journal} {\bibinfo  {journal} {EPJ Quantum Technology}\ }\textbf
  {\bibinfo {volume} {1}},\ \bibinfo {pages} {2} (\bibinfo {year}
  {2014})}\BibitemShut {NoStop}%
\bibitem [{\citenamefont {Erickson}\ and\ \citenamefont
  {Pappas}(2017)}]{erickson2017theory}%
  \BibitemOpen
  \bibfield  {author} {\bibinfo {author} {\bibfnamefont {R.~P.}\ \bibnamefont
  {Erickson}}\ and\ \bibinfo {author} {\bibfnamefont {D.~P.}\ \bibnamefont
  {Pappas}},\ }\bibfield  {title} {\enquote {\bibinfo {title} {Theory of
  multiwave mixing within the superconducting kinetic-inductance traveling-wave
  amplifier},}\ }\href@noop {} {\bibfield  {journal} {\bibinfo  {journal}
  {Physical Review B}\ }\textbf {\bibinfo {volume} {95}},\ \bibinfo {pages}
  {104506} (\bibinfo {year} {2017})}\BibitemShut {NoStop}%
\bibitem [{\citenamefont {Stolen}\ and\ \citenamefont
  {Bjorkholm}(1982)}]{stolen1982parametric}%
  \BibitemOpen
  \bibfield  {author} {\bibinfo {author} {\bibfnamefont {R.}~\bibnamefont
  {Stolen}}\ and\ \bibinfo {author} {\bibfnamefont {J.}~\bibnamefont
  {Bjorkholm}},\ }\bibfield  {title} {\enquote {\bibinfo {title} {Parametric
  amplification and frequency conversion in optical fibers},}\ }\href@noop {}
  {\bibfield  {journal} {\bibinfo  {journal} {IEEE Journal of Quantum
  Electronics}\ }\textbf {\bibinfo {volume} {18}},\ \bibinfo {pages}
  {1062--1072} (\bibinfo {year} {1982})}\BibitemShut {NoStop}%
\bibitem [{Note1()}]{Note1}%
  \BibitemOpen
  \bibinfo {note} {For our device, $G_s\simeq 10 dB$ over a $\sim 1.5 GHz$
  bandwidth around the pump depending on the frequency.}\BibitemShut {Stop}%
\bibitem [{Note2()}]{Note2}%
  \BibitemOpen
  \bibinfo {note} {The uncertainty arises from the losses which cannot be
  localized to certain temperature.}\BibitemShut {Stop}%
\bibitem [{\citenamefont {Castellanos-Beltran}\ \emph
  {et~al.}(2008)\citenamefont {Castellanos-Beltran}, \citenamefont {Irwin},
  \citenamefont {Hilton}, \citenamefont {Vale},\ and\ \citenamefont
  {Lehnert}}]{castellanos2008amplification}%
  \BibitemOpen
  \bibfield  {author} {\bibinfo {author} {\bibfnamefont {M.}~\bibnamefont
  {Castellanos-Beltran}}, \bibinfo {author} {\bibfnamefont {K.}~\bibnamefont
  {Irwin}}, \bibinfo {author} {\bibfnamefont {G.}~\bibnamefont {Hilton}},
  \bibinfo {author} {\bibfnamefont {L.}~\bibnamefont {Vale}}, \ and\ \bibinfo
  {author} {\bibfnamefont {K.}~\bibnamefont {Lehnert}},\ }\bibfield  {title}
  {\enquote {\bibinfo {title} {Amplification and squeezing of quantum noise
  with a tunable josephson metamaterial},}\ }\href@noop {} {\bibfield
  {journal} {\bibinfo  {journal} {Nature Physics}\ }\textbf {\bibinfo {volume}
  {4}},\ \bibinfo {pages} {929} (\bibinfo {year} {2008})}\BibitemShut {NoStop}%
\bibitem [{\citenamefont {Zobrist}\ \emph {et~al.}(2019)\citenamefont
  {Zobrist}, \citenamefont {Eom}, \citenamefont {Day}, \citenamefont {Mazin},
  \citenamefont {Meeker}, \citenamefont {Bumble}, \citenamefont {LeDuc},
  \citenamefont {Coiffard}, \citenamefont {Szypryt}, \citenamefont {Fruitwala}
  \emph {et~al.}}]{zobrist2019wide}%
  \BibitemOpen
  \bibfield  {author} {\bibinfo {author} {\bibfnamefont {N.}~\bibnamefont
  {Zobrist}}, \bibinfo {author} {\bibfnamefont {B.~H.}\ \bibnamefont {Eom}},
  \bibinfo {author} {\bibfnamefont {P.}~\bibnamefont {Day}}, \bibinfo {author}
  {\bibfnamefont {B.~A.}\ \bibnamefont {Mazin}}, \bibinfo {author}
  {\bibfnamefont {S.~R.}\ \bibnamefont {Meeker}}, \bibinfo {author}
  {\bibfnamefont {B.}~\bibnamefont {Bumble}}, \bibinfo {author} {\bibfnamefont
  {H.~G.}\ \bibnamefont {LeDuc}}, \bibinfo {author} {\bibfnamefont
  {G.}~\bibnamefont {Coiffard}}, \bibinfo {author} {\bibfnamefont
  {P.}~\bibnamefont {Szypryt}}, \bibinfo {author} {\bibfnamefont
  {N.}~\bibnamefont {Fruitwala}},  \emph {et~al.},\ }\bibfield  {title}
  {\enquote {\bibinfo {title} {Wide-band parametric amplifier readout and
  resolution of optical microwave kinetic inductance detectors},}\ }\href@noop
  {} {\bibfield  {journal} {\bibinfo  {journal} {Applied Physics Letters}\
  }\textbf {\bibinfo {volume} {115}},\ \bibinfo {pages} {042601} (\bibinfo
  {year} {2019})}\BibitemShut {NoStop}%
\bibitem [{\citenamefont {Guha}\ and\ \citenamefont
  {Erkmen}(2009)}]{guha2009gaussian}%
  \BibitemOpen
  \bibfield  {author} {\bibinfo {author} {\bibfnamefont {S.}~\bibnamefont
  {Guha}}\ and\ \bibinfo {author} {\bibfnamefont {B.~I.}\ \bibnamefont
  {Erkmen}},\ }\bibfield  {title} {\enquote {\bibinfo {title} {Gaussian-state
  quantum-illumination receivers for target detection},}\ }\href@noop {}
  {\bibfield  {journal} {\bibinfo  {journal} {Physical Review A}\ }\textbf
  {\bibinfo {volume} {80}},\ \bibinfo {pages} {052310} (\bibinfo {year}
  {2009})}\BibitemShut {NoStop}%
\bibitem [{\citenamefont {Shur}, \citenamefont {Hack},\ and\ \citenamefont
  {Shaw}(1989)}]{shur1989new}%
  \BibitemOpen
  \bibfield  {author} {\bibinfo {author} {\bibfnamefont {M.}~\bibnamefont
  {Shur}}, \bibinfo {author} {\bibfnamefont {M.}~\bibnamefont {Hack}}, \ and\
  \bibinfo {author} {\bibfnamefont {J.~G.}\ \bibnamefont {Shaw}},\ }\bibfield
  {title} {\enquote {\bibinfo {title} {A new analytic model for amorphous
  silicon thin-film transistors},}\ }\href@noop {} {\bibfield  {journal}
  {\bibinfo  {journal} {Journal of applied physics}\ }\textbf {\bibinfo
  {volume} {66}},\ \bibinfo {pages} {3371--3380} (\bibinfo {year}
  {1989})}\BibitemShut {NoStop}%
\bibitem [{\citenamefont {Steck}(2007)}]{steck2007quantum}%
  \BibitemOpen
  \bibfield  {author} {\bibinfo {author} {\bibfnamefont {D.~A.}\ \bibnamefont
  {Steck}},\ }\href@noop {} {\enquote {\bibinfo {title} {Quantum and atom
  optics},}\ } (\bibinfo {year} {2007})\BibitemShut {NoStop}%
\end{thebibliography}%
\end{document}